# Application Distribution Model In Volunteer Computing Environment Using Peer-to-Peer Torrent Like Approach

Yustinus Eko Soelistio
Faculty of Information and Communication Technology
Universitas Multimedia Nusantara
Tangerang, Indonesia
yustinus.eko@umn.ac.id

*Abstract*—Volunteer computing has been known as an alternative solution to solve complex problems. It is acknowledged for its simplicity and its ability to work on multiple operating systems. Nonetheless, setting up a server for volunteer computing can be time consuming and relatively complex to be implemented. This paper offer a model which can ease the effort of setting up a server by making the agent works two ways, as seeder and leecher, like P2P torrent approaches. The model consists of measurement units to manage applications to be distributed, system hierarchy, and basic procedures for the server and the agent. The model has been tested in four scenarios using 2,000,000 to 3,000,000 integer data employing up to six nodes. The tests demonstrate speedup in three of the scenarios.

## I. Introduction

Volunteer computing has been known as one alternative to solve high complexity problems. Its user friendliness for volunteers increases its popularity, especially in solving scientific computations like [1]-[4]. Many improvements have been suggested to improve the functionality of volunteer computing [5]-[14]. Some works focus on increasing its simplicity and availability to volunteers [5]-[9]. One method is by using P2P platform to further extend its capability [8]-[9]. Based on this idea, this paper explores one possibility to make volunteers not only able to crunch tasks from servers, but also able to publish their problem sets easily to another volunteers. The model proposed in this paper employ P2P idea and focus on managing application and data in the cloud.

## II. Related Work

Previous work [9] suggests a volunteer computing P2P agent model to distribute and work on task shared in the cloud. The agent can host and publish a task which will be run by another agent. These pull and push communications happened completely on P2P network overlay thus overcome the restrictions imposed by firewalls and NA(P)T. It provides an easy to implement volunteer computing server model.

The idea from this work will be further explored in this paper. The model will introduce measurement units to inform volunteers about jobs being offered, connectionism model, system hierarchy, and basic communication procedures between agents. In the future studies, the model can be adapted to facilitate hosts running on coarse grain connectionism by employing mirroring jobs and splitting databases to clients.

## III. Model

The model starts by defining that all nodes ($O$) in this P2P torrent cloud send information of its available application ($A$) to be shared, and receive information regarding available applications and their descriptions. For every $A$ in $O$, each $O$ can acts either as a seeder ($O_s$) when it offers $A$, or a leecher ($O_l$) when works on $A$.

### A. Applications Management

Unlike processes in P2P torrent such as BitTorrent which progresses can be easily measured in terms of sizes, measuring application complexity in volunteer computing can be tricky. Complexity in volunteer computing mostly measured by how many flops needed to complete one application with a set of data. This is troublesome since we need to go through the code and algorithm to find out flops requirement for a particular application. Furthermore it is not practically doable in the torrent like environment with thousands of applications waiting to be run. Therefore this paper proposes using multiple units of measurement to assess applications' complexity. These units are data size, popularity, and average working time. These units are saved in servers and the information will be made public to all volunteers.

### B. Measurement Model

Data size, popularity, and average working time are units to measure the goodness of an application in the proposed volunteer computing torrent cloud. The net values of these units will be published as consideration tools for volunteers to whether run the application or not.

Data size ($d$) is the sum of application's size ($d_{app}$) and its data's size ($d_{data}$) which processed by application $A$. Each $A$ can have multiple $d_{data}$ thus

$$d_A = \sum_0^y d_{app} + \sum_0^i d_{data} \quad (1)$$

Popularity ($p$) is the number of nodes work on $A$. If $A$ has been running for multiple time then

$$p_A = \sum_1^i frequency\ A_i \quad (2)$$

Average working time ($w$) is the average time needed for $O$ to finish working on $A$. If $A$ has been running for more than once then

$$w_A = \frac{\sum_1^i time\ A_i}{p_A} \quad (3)$$

Units $p$ and $w$ are unrelated to size to better provide comprehensive information to volunteers regarding the applications. To approximate complexity of $A$, volunteers can use units $d$ and $w$. In general, combination of high $d$ and low $w$ can indicate low complexity. Moreover, high value of $p$ and $w$ with low $d$ can suggest high complexity.

*C. System Model*

This volunteer computing model adopts structure like hybrid P2P model in [15]. All volunteers run an agent that connect to tracking server. When succeed, the agent send information about its offered application $A_{self}$. If it has multiple $A_{self}$ then it send a list of $A_{self(i)}$ to the server. In return, server will send information regarding available $A$ in the cloud, including their host identities. Fig. 1 illustrates the model's information flow.

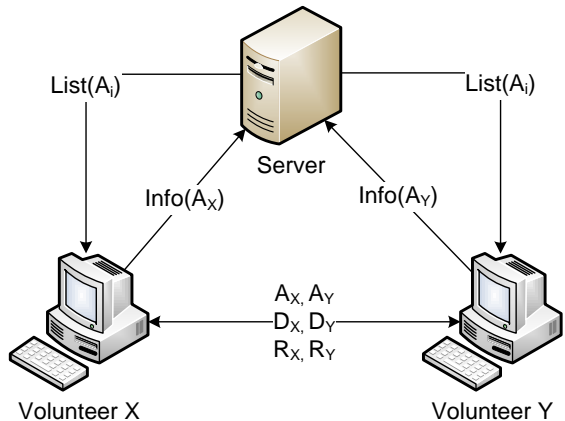

Fig. 1. Volunteer computing P2P model's information flow

When a volunteer choose to run another volunteer's application, it directly contact the application host. For example in Fig. 1, if volunteer X wishes to work on $A_Y$, then it will connect to volunteer Y to get the application and data $D_Y$ (if available). Volunteer X will work on $A_Y$ and will determine values of $d$ and $w$. When completed, volunteer X send result $R_Y$ along with its $d$ and $w$. Volunteer Y will then validate the result. If $R_Y$ is valid then volunteer Y will notify server about the work status including its $d$ and $w$. Finally, the server will update and publish the result back to the cloud.

*D. Validation Model*

Tracking server maintains list of $A_i$ by updating its timestamp. Application $A_Y$ will be preserved in the list only if volunteer Y periodically updates its status to the server. When tracking server does not receive update status from volunteer Y in a certain amount of time ($t$) for a maximum number of $f$ times, then information about $A_Y$ will be removed from the list. Since volunteer X's works are based on the list, then it will discard $A_Y$, $D_Y$, and $R_Y$.

Result $R_Y$ from volunteer X will be validated by volunteer Y as the host. This model adapts validation technique like [5] that introduced majority voting to determine whether $R_Y$ is valid. The maximum and minimum number of node to validate are denoted by $m_{max} \geq m$. For volunteer Y to verify that $R_Y$ is valid, it needs at least $m_{min}$ nodes working on $A_Y$. Consequently (1), (2), and (3) become

$$\begin{cases} d = m_{min}\left(\sum_o^y d_{app} + \sum_0^i d_{data}\right) \\ p = m_{min}\left(\sum_1^i frequency\ A_i\right) \\ w = m_{min}\left(\sum_1^i time\ A_i \Big/ p\right) \end{cases} \quad (4)$$

Any malicious $R_Y$ will be discarded and its status will not be updated by the server.

*E. Server and Agent Model*

Tracking server consists of three modules which are connection module, tracker module, and synchronizer module. Connection module receives and sends messages to volunteers for any updates on applications list. Tracker module is responsible for checking applications and their host availability. Tracker verifies host availability by sending periodic message to hosts through connection module.

Applications are verified base on information from volunteers. Updates on valid hosts and applications from tracker will be saved to applications list by the synchronizer. Afterward, this list will be published to volunteers by the connector. Fig. 2 shows relations between modules.

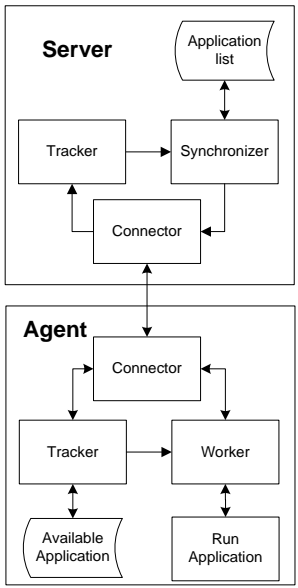

Fig. 2. Server-agent modules model.

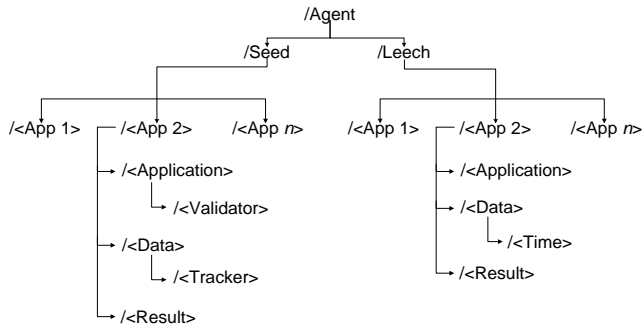

Fig. 3. Agent's directory structure

Similar to the server, an agent contains a connector and tracker to manage connection to the server and application host. In addition, an agent has a worker that run application $A$. The connector serves two ways; first as an application host it informs any new information on $A_{self}$ (including updated status on $d(A_{self})$ and $w(A_{self})$) to the server, send $A_{self}$ and $D_A$ to volunteers, and confirms availability check from the server. Second, as a volunteer it receives $A$ and $D$ from $O(A)$ and sends $R$ to $O(A)$. The tracker also acts as host and volunteer. As host, tracker distributes and manages $A_{self}$ and $D_A$ to volunteers via connector. As volunteer, tracker tracks any change on the applications list and inform connector or worker about the changes. For example, when tracker find out that application host has drop its $A_{self}$ then tracker will immediately notify the worker to drop the application as well.

*F. Directory Structure*

The agent's working directory adopts [16] with some modifications. It works on two separate directories, seed directory and leech directory. Seed directory is maintained by the tracker. It contains all applications $A_{self}$, problem data, and their results. Leech directory contains all other hosts applications, data, and results that the worker worked on. All files in leech directory are temporary. Once an application is finished, the directory in which the application worked on will be removed after the result is sent to its host. Fig. 3 shows directory structure of the agent.

*G. Procedures Model*

Tracker server has three communication procedures, three tracking procedures, and two synchronizing procedures. Communication procedures are PING for availability check, PUSH to send applications list, and RECV to collect messages from volunteers. Returned results from PING and RECV will be passed to tracker. Tracking procedures consist of VAL to validate the availability of application hosts, INIT to send initial applications list to new volunteers, and INFO to inform synchronizer about availability, information update, or changes in application hosts. Synchronizing procedures consists of WRITE to update applications list and READ to read applications list.

PING has two parameters, $t$ and $f$ like mentioned in chapter II.D. RECV can treat a list of blocked clients as parameter. By default VAL receives status, either true or false, from PING and messages from RECV. VAL will tell INFO to update application list if the status it received is true. VAL can be customized to call another function to complement PING. For instance, it can put a client that has low availability into a black list so any message from this client will be rejected and all applications from this client will be dropped. If VAL receives initialization message from RECV then VAL will call INIT to push application list to the client. INIT stores temporary applications list which will be updated periodically based on a timer. When it reaches its timeout, INIT will request a READ from synchronizer to get the latest list. To update the list with new information, VAL passes the messages from RECV to WRITE.

Agent has two communication procedures, five tracking procedures, and eight working procedures. Communication procedures consist of RECV to receive messages from server and result from volunteers, and SEND to send messages to server and application hosts. Tracking procedures consists of EVAL to keep track number of $m_{max}$ and $m_{min}$, DIST to distribute applications and data trough communicator, STAT to update valid works information to server, VAL to check whether the results received are valid, and TAIL to keep track of volunteers including work timeout. When results are valid then VAL will inform EVAL to increase $m_{min}$. TAIL tracks volunteers and their works by saving and checking the log in /Seed/App/Data/Tracker directory. Working procedures consists of REQ to request application and data from application host, SCAN to scan the size of application and data,

RUN to start running the application, TIME to evaluate working time of each application, COLLECT to collect information about running applications from TIME and SCAN, SAVE to write the result in /result, LOAD to read back the result, and STOP to drop an application (including its data and result). TIME keeps track of working time by saving and updating the log in /Leech/App/Data/Time.

In agent, RECV has parameter that can be set to accept or deny messages from certain clients. VAL can be customized just like VAL in the server so it can call another function. This function can be a checklist or another application that returns Boolean status. TAIL maps application and the volunteer working on it. TAIL has timeout parameter which is the due time for volunteers to submit their results. If it reaches its timeout before the volunteer submit the result, then TAIL will call DIST and drop the volunteer ID from the mapping list. When a client accepts to run a certain application from the list, the agent will call REQ to request application and data. After the download completes, REQ will call SCAN and RUN all together. RUN will call TIME which mark the beginning and the end of the work. When finished, COLLECT and LOAD will be invoked to get time, size and the result. Afterwards, data from COLLECT and LOAD will be returned to the host by SEND. If SEND fails to submit the result then it will inform STAT to update application list from the server. If the application host is not on the list, then by default STAT will immediately call STOP. Fig. 4 and Fig. 5 illustrate relation between process. Fig. 4 shows process for managing application hosts. Fig. 5 shows process for distributing applications.

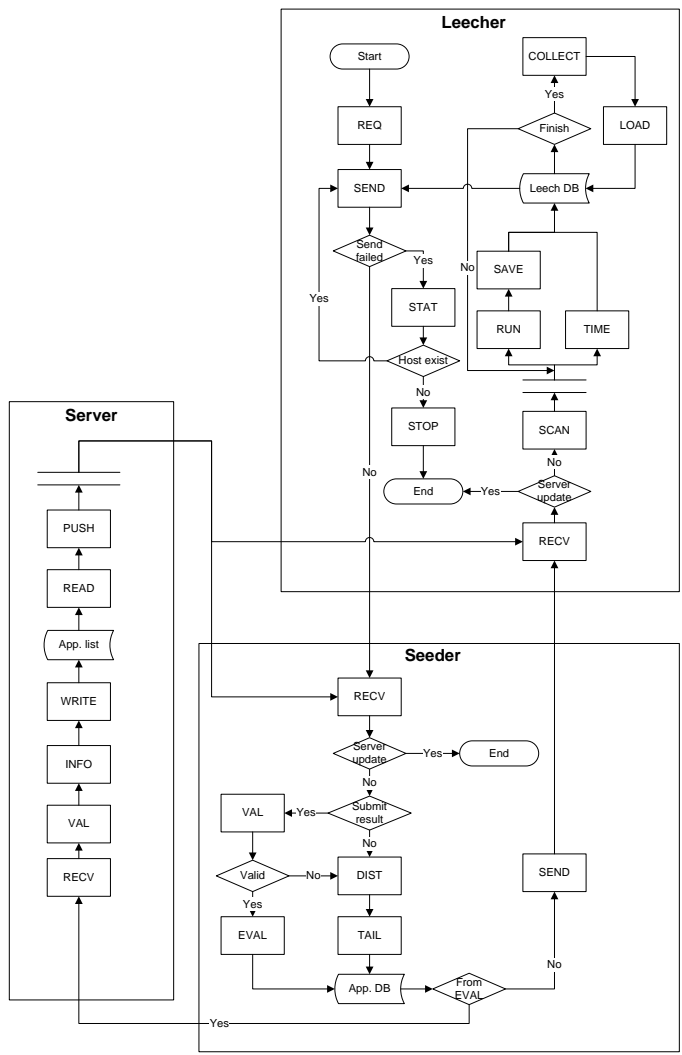

Fig. 5. Application distribution process

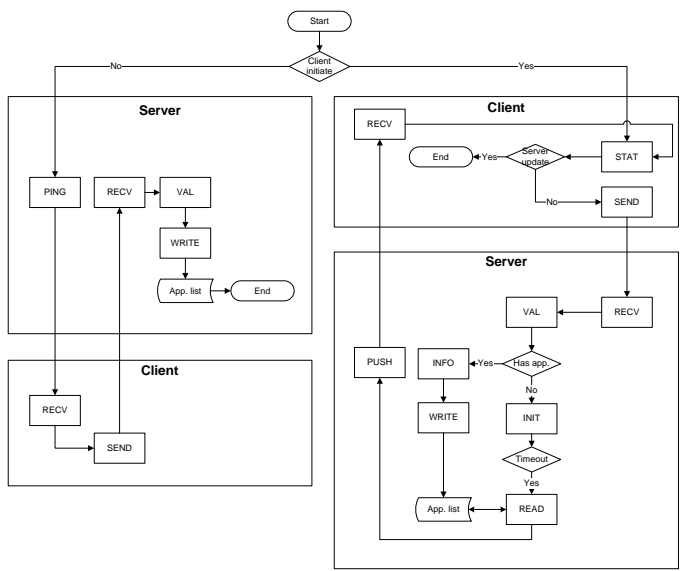

Fig. 4. Process for managing application hosts

## IV. TESTING

### A. Scenario I: three volunteers with one application

The test used four nodes connected to each other. One node acted as the server, and the others as clients. One of the clients ($O_s$) had an application that tried to find integer prime numbers range from 3 to 2,000,000 using exhaustion method. The problem was divided into 2059 parts. Each part contained a set of minimum and maximum numbers to be identified as prime number. These parts were then distributed to two other clients $O_{l(X)}$ and $O_{l(Y)}$ thus each shared approximately 1030.

For each REQ procedure $O_{l(X)}$ and $O_{l(Y)}$ sent to $O_s$, $O_{l(X)}$ and $O_{l(Y)}$ received application file and theirs share of data. $d_{app} \approx 4 kB$ and was constant through all cycles since all cycles run the same application file. Total $d_{data}$ for all 2059 parts is ~*8.33 MB*. Since the application file was downloaded on each cycle then it was estimated that each $O_l$ would received total *8.28 MB* of data.

Validation in this test did not precisely follow majority voting method like described in chapter II.D since the test did not incorporate malicious volunteer. Instead, the test put $m_{max}=m_{min}=1$ so each $O_l$ repeatedly run REQ and RUN procedures until the application successfully processed all numbers in range.

The application was built using Python. The test was conducted on one computer and three virtual machines. All clients were run on virtual machines, and the server was run on their host. The host was run on Intel i5 with 4 GB RAM. The virtual machines configuration were one core with 524 MB RAM. Running time from the test was compared to the one from sequential process. Sequential process was run twice, once on one of the clients and another on the host machine. Table I shows the result of first test. The test shows speedup about 1.56 against the host and about 1.73 against the virtual machine. Number of cycle done and data received ($d$) on each $O_l$ were relatively close to the approximation. Since $d_{app}$ is the sum of all application files and its data then $d_{app}=16.56$ MB. The $p$ and $w$ for this application are 2059 and 6.35 second.

TABLE I. TEST RESULT COMPARISON FROM SCENARIO I

|  | Seq. host | Seq. virtual | Parallel client 1 | Parallel client 2 |
|---|---|---|---|---|
| # of cycle | 2059 | 2059 | 1031 | 1028 |
| Time (hour) | 2.82 | 3.15 | 1.82 | 1.81 |
| Average time-per-cycle (s) | 4.93 | 5.51 | 6.36 | 6.35 |
| Data size (MB) | 8.32 | 8.32 | 8.29 | 8.27 |

*B. Scenario II: three volunteers with two applications*

The second test added one extra application into the cloud. There were two clients that acted as leecher and seeder, and one client acted as leecher to both applications (see Fig. 4 for more detail). The second application tried to find integer prime number from 2,000,000 to 3,000,000 which divided into 1080 parts. This time, the test shows slowdown for the first application and speedup for the second application. However, in general the time needed to complete both applications is faster by about 33%. Table II shows the results' comparison between applications and between hosts.

TABLE II. RESULT COMPARISON FROM EXPERIMENT II

|  |  | # of cycle | Time (h) | Avg. (s) | Size (MB) |
|---|---|---|---|---|---|
| Sequential app. 1 | Host | 2059 | 2.82 | 4.93 | 8.32 |
|  | VM | 2059 | 3.15 | 5.51 | 8.32 |
| Sequential app. 2 | Host | 1080 | 6.78 | 21.21 | 4.23 |
|  | VM | 1080 | 6.73 | 21.66 | 4.23 |
| Parallel app. 1 | Client Y | 139 | 0.35 | 9.09 | 1.12 |
|  | Client Z | 1920 | 4.45 | 8.34 | 15.45 |
| Parallel app. 2 | Client Y | 462 | 4.33 | 33.57 | 3.66 |
|  | Client X | 618 | 4.48 | 33.56 | 4.89 |

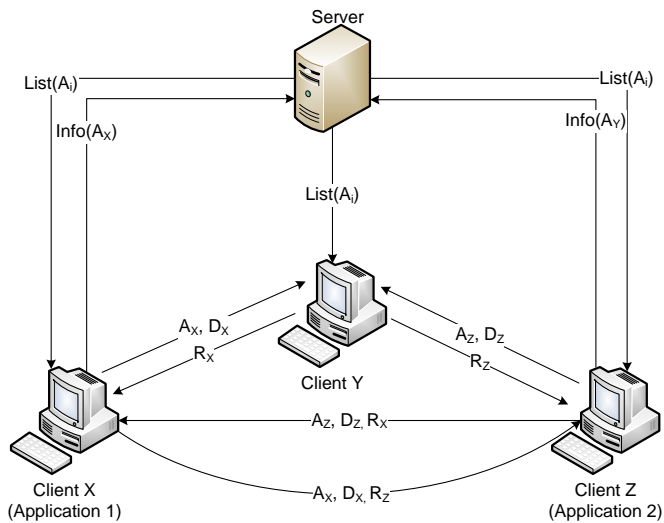

Fig. 6. Test with three volunteers and two applications information flow

*C. Scenario III: scenario II with two additional leeches*

This test used the same configuration as the second test with some adjustment. Aside from hosting $A_X$ and $A_Z$, client X and client Z also run their own application. Table III shows result from this test. Compared to the previous test's result in table II, this test affirmed speedup for application 2. Although did not show an obvious speedup, application 1 in this test showed shorter run time then the one in the previous test.

TABLE III. TEST III RESULT

| Parallel | Client | # of cycle | Time (h) | Avg. (s) | Size |
|---|---|---|---|---|---|
| App. 1 | X | 736 | 1.54 | 7.54 | 5.92 |
|  | Y | 635 | 2.88 | 9.14 | 5.11 |
|  | Z | 688 | 1.63 | 8.42 | 5.53 |
| App. 2 | X | 401 | 3.50 | 31.41 | 3.17 |
|  | Y | 329 | 3.40 | 37.24 | 2.61 |
|  | Z | 350 | 3.41 | 35.04 | 2.77 |

*D. Scenario IV: six volunteers with two applications*

The fourth test added three more clients into the cloud. Unlike previous test, the last three clients were another computer (run on Intel i3 with 2GB RAM) and its two virtual machines. This new host was connected to the first host over 100BASE-TX fast Ethernet connection. All of these new clients run both $A_X$ and $A_Y$ just like what client Y did in Fig. 4. For simplicity these new clients were called $X'$, $Y'$, and $Z'$ for their host.

Table IV shows faster run time then the one on table II. The speedup can reach about 3.5 for application 1 and 3.3 for application 2. Though only reach about half of the ideal linear speedup, the test verifies that the model provides better performance then the one performed on single core. Data size received by each client is smaller than the one received in previous tests. Keeping the data small can be important in volunteer computing environment since not all volunteers have a decent internet connection.

TABLE IV. TEST IV RESULT

| Parallel | Client | # of cycle | Time (h) | Avg. (s) | Size |
|---|---|---|---|---|---|
| App. 1 | X | 387 | 0.84 | 7.84 | 3.11 |
| | Y | 373 | 0.89 | 8.64 | 3.00 |
| | Z | 380 | 0.84 | 7.99 | 3.06 |
| | X | 290 | 0.87 | 10.79 | 2.33 |
| | Y | 289 | 0.87 | 10.85 | 2.32 |
| | Z | 340 | 0.86 | 9.12 | 2.74 |
| App. 2 | X | 194 | 1.88 | 34.97 | 1.45 |
| | Y | 196 | 1.88 | 34.62 | 1.47 |
| | Z | 207 | 1.94 | 33.69 | 1.55 |
| | X | 147 | 1.90 | 46.56 | 1.10 |
| | Y | 149 | 1.90 | 45.96 | 1.12 |
| | Z | 187 | 1.88 | 36.23 | 1.40 |

## V. CONCLUSION

This paper explores one possibility to manage applications in volunteer computing environment. The model is designed to work under P2P torrent like cloud. The model has been tested under different situations which are likely to happen in P2P torrent cloud. Nearly all tests prove some degree of speedup. Even so, further validations are still needed. Further test should include a more comprehensive validation method to verify its robustness against malicious volunteers and glitch in internet connectionism.

The model proposed in this paper can be implemented using 23 basic procedures. Each procedure has some parameters that can be modified to fit volunteer's requirements. Host's information in the application list can be developed to hold multiple hosts ID, thus allowing the applications to be mirrored or to be broken to pieces like regular file sharing in torrent.

Next study can explore other measurement units to better fit the model, improve the basic procedures, and enhance security including ways to confine seeder's applications from leecher's system.